\documentclass[twocolumn,prd,nofootinbib,aps,prl,floats,floatfix,amsmath,amssymb,longbibliography,secnumarabic]{revtex4-1} %
\usepackage[english]{babel}
\usepackage{csquotes}
\usepackage{csquotes}

\usepackage[letterpaper,top=2cm,bottom=2cm,left=3cm,right=3cm,marginparwidth=1.75cm]{geometry}

\usepackage{amsmath}
\usepackage{graphicx,slashed}
\usepackage[colorlinks=true,citecolor=red]{hyperref}

\newcommand{\be}{\begin{equation}}
\newcommand{\ee}{\end{equation}}

\newcommand{\bea}{\begin{eqnarray}}
\newcommand{\eea}{\end{eqnarray}}

\begin{document}

\title{``Neutrinoless double beta decay'' is the correct name\\ for neutrinoless double beta decay}
\author{James M.\ Cline}
\email{jcline@physics.mcgill.ca}
\affiliation{McGill University Department of Physics \& Trottier Space Institute, 3600 Rue University, Montr\'eal, QC, H3A 2T8, Canada}

\begin{abstract}
Recently \url{https://arxiv.org/pdf/2604.12897} urged that the terminology ``neutrinoless double beta decay'' should be changed to ``Majorana double beta decay'' to properly give credit to Majorana, and to focus on the positive aspects of the phenomenon---supposed creation of matter in the laboratory---rather than the negative: absence of something, embarrassment over false claims of detection, and a ``sociology of suspicion.''  I argue that the current terminology is more accurate and descriptive, and that the claimed reasons for its adoption are lacking in credibility. 

\end{abstract}

\maketitle

Ref.\ \cite{Vissani:2026mdr} has advocated that the physics community abandon the terminology ``neutrinoless double beta decay'' and instead refer to the phenomenon as ``Majorana double beta decay.''  It is argued that this was the original terminology, which was linked to its theoretical origins, and that the shift toward a ``semantically bleached neologism'' was a consequence of several negative sociological influences.  Supposedly, the community had been embarrassed by claims of detection that turned out to be false, and for unexplained reasons, they felt exonerated by moving to a more neutral terminology.  No direct evidence is given for this interpretation of the events. 

Nor is any evidence given for the claim that the process was widely known as ``Majorana double beta decay'' in its early stages of conceptual development.  Not Majorana, but rather Furry \cite{Furry:1939qr} was the first to recognize its possibility.  Hence, it should be known as ``Furry double beta decay'' if we wanted to give credit where it is due.   Instead, Majorana had shown in his celebrated paper \cite{Majorana:1937vz} that the theory of ordinary beta decay is practically unaltered by postulating that neutrinos are their own antiparticles.  To call the $0\nu\beta\beta$ process ``Majorana double beta decay'' would not only misattribute Furry's contribution, it would evoke the incorrect picture in which ordinary double beta decay is accompanied by two Majorana neutrinos.

J.\ McCarthy's paper \cite{PhysRev.90.853}, which claimed experimental evidence for $0\nu\beta\beta$, is criticized for not referring to Dirac or Majorana, but rather to Furry and  Goeppert-Mayer \cite{PhysRev.48.512} (who elucidated ordinary double beta decay), and for introducing the current terminology.  We must place this in historical context: in those days it was customary to cite only the dozen or so papers that were directly relevant and which one had read.   An experimentalist was not obliged to reference people who had influenced (but not proposed) the theories that were of immediate importance to the experiment.  The absence of neutrinos is the most physically descriptive fact that distinguishes $0\nu\beta\beta$ from $2\nu\beta\beta$, rather than the inventor of the kind of neutrino mass term that leads to the difference.

Ref.\ \cite{Vissani:2026mdr} further laments that the $0\nu\beta\beta$ process is known for what it does not produce, rather than for what it does, namely ``matter creation in the laboratory.''  This seems somewhat exaggerated, reminiscent of the hyperbole often used by the $0\nu\beta\beta$ community to enlarge the consequences of their work by linking it to leptogenesis.    The only sense in which matter is created in $0\nu\beta\beta$ is that two electrons appear in the decay, as opposed to two electrons plus two antineutrinos.  Does production of antineutrinos count as negative matter creation, thus making $0\nu\beta\beta$ by comparison a net creation of matter?  Since the low-energy antineutrinos produced in $2\nu\beta\beta$ are essentially inert and invisible, it is difficult to justify that matter has been created in any significant sense in $0\nu\beta\beta$.

Those who wish to be sure that Majorana is properly credited need not worry.  Every school child (in the school of elementary particle physics) learns that the observation of neutrinoless double beta decay would be evidence for the Majorana nature of neutrinos.  Whether it would also constitute evidence for how matter was created in the early Universe is debatable.

\bibliographystyle{utphys}
\bibliography{sample}

\end{document}